# Is $E = mc^2$ an exclusively relativistic result?

Qing-Ping Ma

The Department of Quantitative and Applied Economics, the University of Nottingham Ningbo China

**Abstract**

The mass-energy formula $E = mc^2$ is thought to be derived by Einstein from special relativity. The present study shows that since the formula has also been derived from classical physics by Einstein, it is not an exclusively relativistic result. The formula is implied by Maxwell's electromagnetic momentum $P = E/c$ and the Newtonian definition of momentum $P = mv$. $E = mc^2$ like momentum $P = mv$ applies to both classical physics and special relativity, if relativistic mass is used in the equation. Einstein's derivation in 1905 is logically flawed as a relativistic proof and the truly relativistic formula should be $E = E_0/\sqrt{1 - v^2/c^2}$ derived by Laue and Klein. If the energy measured in one reference frame is $E_0$, it is $E = E_0/\sqrt{1 - v^2/c^2}$ in a reference frame moving at velocity $v$ relative to the first frame.

**Key words**: Lorentz transformation; Mass-energy equation; Special relativity; conservation of momentum; conservation of energy; reference frame.

## 1. Introduction

The mass-energy formula $E = mc^2$ has a prominent role in both physics research and public perception of science. The formula explains the power of nuclear bombs as well as the energy source of stars [1-3], and stimulates the imagination of general public. It also underlies key components of the Dirac equation, which has accounted for the fine details of the hydrogen spectrum and implied the existence of antimatter [4]. Although Einstein in 1905 derived mass-energy equivalence initially as an approximation [5], the accuracy of the formula has been confirmed by experiments to a high level of precision [6].

The explicit expression of $E = mc^2$ was first proposed by Planck [7-9], but it is generally believed that Einstein derived the mass-energy formula $E = mc^2$ from special relativity in 1905. Fernflores asserts in Stanford Encyclopedia of Philosophy: 'Einstein correctly described the equivalence of mass and energy as "the most important upshot of the special theory of relativity" [10], for this result lies at the core of modern physics' [11]. Although there are still some disputes on Einstein's discovery of the mass-energy equation and some researchers have argued that Einstein's derivation might be logically flawed [12-16], nobody seems to question whether the mass-energy equation is really a relativistic result.

It has been known that the mass-energy equation appears to be implied in Maxwell's electromagnetic theory [17-19], and Lewis [20] has provided a derivation within the framework of classical physics. Since the mass-energy equation might be derived within the framework of classical physics, it could be a result from classical physics rather than special relativity. The aim of this study is to show that $E = mc^2$ is actually a formula common to both classical physics and special relativity, and the relevant relativistic formula is $E = E_0 / \sqrt{1 - v^2/c^2}$. This study will prove this by examining Einstein's first derivation of mass-energy relation in 1905 and his last derivation in 1946 and providing logically more consistent corresponding derivations.

It must be emphasized here that, this study does not question the validity of the mass-energy equation, nor does it question the validity of special relativity. The main fact this study intends to establish is that, the mass-energy equation has a status similar to that of the conservation of momentum rather than that of time dilation or length contraction. The mass-energy equation and the conservation of momentum are valid in both classical physics and special relativity; therefore, they are not relativistic conclusions. Time dilation and length contraction are not compatible with classical physics, hence they are relativistic.

## 2. Criteria for being relativistic

As the present study intends to argue that the mass-energy formula is common to both classical physics and special relativity, we need to establish the criteria for being relativistic. What qualifies a formula as a relativistic result? The following criterion could be used:

**Proposition 1**. A formula is relativistic if and only if the formula in its general form or specific forms can be derived only when assumptions or results unique to special relativity have been applied.

With this criterion, we can readily tell whether a formula or physical law is relativistic or not. Many laws in physics are valid in both classical physics and special relativity, but we cannot say those laws are consequences of special relativity simply because they are valid in special relativity. For example, the Newton's third law and the conservation of momentum are still valid in special relativity, but they are not relativistic results or conclusions. Some conclusions in physics are not valid in classical physics or compatible with it, such as time dilation and length contraction, so that they are relativistic results. Although the concept of relativistic mass has been dismissed by many physicists [21], it is obviously not a concept in classical physics.

Proposition 1 treats the necessity of using uniquely relativistic assumptions or results to derive a formula as a basic criterion for it to be relativistic. If the derivation of a formula must use a uniquely classical assumption or result, can it be relativistic formula? The following criterion could be used as an answer for this question.

**Proposition 2.** If the derivation of a formula must use a result or assumption unique to classical physics, the formula cannot be viewed as relativistic.

Proposition 2 puts a more restrictive constraint on what can be considered being relativistic. Some researchers may argue that special relativity contains classical physics, so using classical physics to derive a formula does not affect its relativistic nature. However, if a formula can only be derived under some conditions unique to classical physics (although they are low speed approximations of relativistic conditions), it cannot be extended to higher speed scenarios, so that it is not relativistic.

The mass-energy equation is about the equivalence between mass and energy, but to energy measured in which reference frame is a mass measured in one reference frame equivalent? This question is an important one, because it puts a constraint on the validity of derivations of the mass-energy equation. Is an object's mass measured in reference frame

A equivalent to its energy measured in the same reference frame (i.e. A), or its energy measured in another reference frame? To my knowledge, this question has not been raised or discussed so far. The following restriction might be imposed with respect to this question:

**Proposition 3**. In the mass-energy equation $E = mc^2$, energy $E$ and mass $m$ are measured in the same reference frame rather than different reference frames.

Proposition 3 requires us to keep track of the reference frames involved in measuring mass and energy during a derivation. Obviously, an object's mass $m$ measured in one reference frame (e.g. frame A) cannot have the same mass-energy relationship $E = mc^2$ with values of its energy $E$ measured in all reference frames, i.e. $E_{\text{any reference frame}} = m_A c^2$ is incorrect, since the values of $E$ measured in other reference frames depend on their velocities relative to frame A.

In classical physics, the issue of different reference frames is less noticeable, because at low velocity the variations of an object's total energy in different reference frames due to kinetic energy differences between different reference frames are negligible compared with the energy implied by its rest mass. In special relativity, an object's kinetic energy in some reference frames can be much larger than the energy implied by its rest mass, so identifying the reference frames where mass and energy are measured is essential for valid derivation of mass-energy relationships. If the derivation gives the equivalence between mass in frame A and energy in frame B in the form of $E = mc^2$ while the two frames move relative to each other, we know it is unlikely to be a correct derivation.

## 3. Einstein's non-relativistic derivation of mass-energy formula in 1946

Einstein gave his last derivation of the mass-energy equivalence in 1946 [22], which is based on conservation of momentum and Maxwell's classical theory of electromagnetism. Since the derivation is quite short, its key part is quoted here (Fig.1).

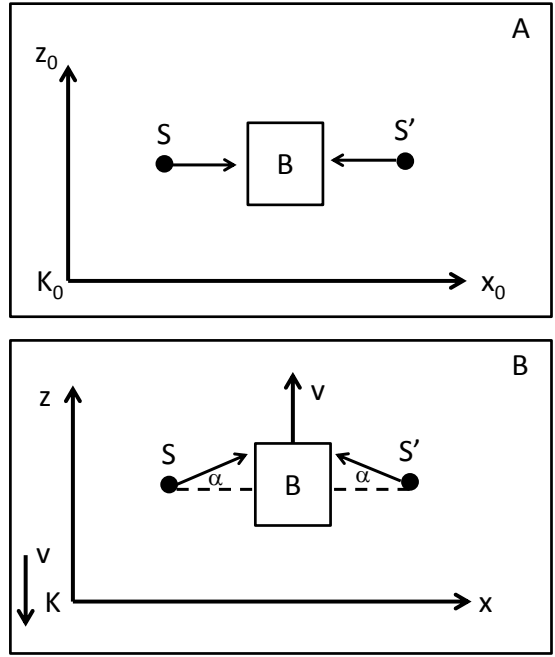

Fig.1. An object *B* absorbing two wave complexes (*S* and *S'*) from opposite directions with energy *E*/2 each. A. Object *B* is at rest in frame $K_0$. B. In frame *K* which moves along *z*-axis negative direction of frame $K_0$ with velocity *v*, object *B* is moving in the **z**-axis positive direction with velocity *v*, and the two wave complexes have an angle α with the *x*-axis, $\sin\alpha = v/c$.

"We now consider the following system. Let the body *B* rest freely in space with respect to the system $K_0$. Two complexes of radiation *S*, *S'* each of energy *E*/2 move in the positive and negative $x_0$ direction respectively and are eventually absorbed by *B*. With this absorption the energy of *B* increases by *E*. The body *B* stays at rest with respect to $K_0$ by reasons of symmetry. Now we consider this same process with respect to the system *K*, which moves with respect to $K_0$ with the constant velocity *v* in the negative $Z_0$ direction. With respect to *K* the description of the process is as follows:

The body *B* moves in positive *Z* direction with velocity *v*. The two complexes of radiation now have directions with respect to *K* which make an angle *α* with the *x* axis. The law of aberration states that in the first approximation $\alpha = \dfrac{v}{c}$, where *c* is the velocity of light. From the consideration with respect to $K_0$ we know that the velocity $v$ of B remains unchanged by the absorption of $S$ and $S'$.

Now we apply the law of conservation of momentum with respect to the $z$ direction to our system in the coordinate-frame $K$.

I. Before the absorption let $m$ be the mass of $B$; $mv$ is then the expression of the momentum $B$ (according to classical mechanics). Each of the complexes has the energy $E/2$ and hence, by a well-known conclusion of Maxwell's theory, it has the momentum $\dfrac{E}{2c}$. Rigorously speaking this is the momentum of $S$ with respect to $K_0$. However, when $v$ is small with respect to $c$, the momentum with respect to $K$ is the same except for a quantity of second order of magnitude ($\dfrac{v^2}{c^2}$ compared to 1). The $z$-component of this momentum is $\dfrac{E}{2c}\sin\alpha$ or with sufficient accuracy (except for quantities of higher order of magnitude) $\dfrac{E}{2c}\alpha$ or $\dfrac{E}{2}\cdot\dfrac{v}{c^2}$. $S$ and $S'$ together therefore have a momentum $E\dfrac{v}{c^2}$ in the $z$ direction. The total momentum of the system before absorption is therefore

$$mv + \dfrac{E}{c^2}\cdot v. \qquad [(1)]$$

II. After the absorption let $m'$ be the mass of $B$. We anticipate here the possibility that the mass increased with the absorption of the energy $E$ (this is necessary so that the final result of our consideration be consistent). The momentum of the system after absorption is then

$m'v$

We now assume the law of the conservation of momentum and apply it with respect to the $z$ direction. This gives the equation

$$mv + \frac{E}{c^2} \cdot v = m'v.  \qquad [(2a)]$$

or

$$m' - m = \frac{E}{c^2}. \qquad [(2b)]$$

This equation expresses the law of the equivalence of energy and mass. The energy increase $E$ is connected with the mass increase $\frac{E}{c^2}$. Since energy according to the usual definition leaves an additive constant free, we may choose the latter that

$$E = mc^2." \qquad (3)$$

There is no special relativity involved in Einstein's derivation in 1946, which is a demonstration that derivation of $E = mc^2$ does not require special relativity. Using Maxwell's theory of electromagnetism and conservation of momentum, Lewis also derived $E = mc^2$ in 1908 [20]. Poincaré implicitly derived the mass-energy relation from classical physics in 1900 [19]. Since neither Einstein's derivation in 1946 nor Lewis' derivation in 1908 requires assumptions unique to special relativity, according to our Proposition 1, the mass-energy formula $E = mc^2$ is not a result of special relativity.

## 4. Einstein's derivation in 1905 and its flaws as a relativistic proof

It is Einstein's first derivation in 1905 that links the mass-energy equation with special relativity [5]. The derivation is based on a thought experiment that is unlikely to be achievable in laboratory [14, 15]. Its key part is quoted here.

"Let a system of plane waves of light, referred to the system of co-ordinates (*x*, *y*, *z*), possess the energy *L*; let the direction of the ray (the wave-normal) make an angle $\phi$ with the axis of *x* of the system. If we introduce a new system of co-ordinates (*ξ*, *η*, *ζ*) moving in uniform parallel translation with respect to the system (*x*, *y*, *z*), and having its origin of co-ordinates in motion along the axis of *x* with the velocity *v*, then this quantity of light—measured in the system (*ξ*, *η*, *ζ*)—possesses the energy

$$L^* = L \frac{1 - \frac{v}{c}\cos\phi}{\sqrt{1 - \frac{v^2}{c^2}}} \qquad [(4)]$$

where $c$ denotes the velocity of light. We shall make use of this result in what follows.

Let there be a stationary body in the system $(x, y, z)$, and let its energy—referred to the system $(x, y, z)$ be $E_0$. Let the energy of the body relative to the system $(\xi, \eta, \zeta)$ moving as above with the velocity $v$, be $H_0$.

Let this body send out, in a direction making an angle $\phi$ with the axis of $x$, plane waves of light, of energy ½$L$ measured relatively to $(x, y, z)$, and simultaneously an equal quantity of light in the opposite direction. Meanwhile the body remains at rest with respect to the system $(x, y, z)$. The principle of energy must apply to this process, and in fact (by the principle of relativity) with respect to both systems of co-ordinates. If we call the energy of the body after the emission of light $E_1$ or $H_1$ respectively, measured relatively to the system $(x, y, z)$ or $(\xi, \eta, \zeta)$ respectively, then by employing the relation given above we obtain

$$E_0 = E_1 + \frac{1}{2}L + \frac{1}{2}L \qquad [(5)]$$

$$H_0 = H_1 + \frac{1}{2}L \frac{1 - \frac{v}{c}\cos\phi}{\sqrt{1 - \frac{v^2}{c^2}}} + \frac{1}{2}L \frac{1 + \frac{v}{c}\cos\phi}{\sqrt{1 - \frac{v^2}{c^2}}} = H_1 + \frac{L}{\sqrt{1 - \frac{v^2}{c^2}}} \qquad [(6)]$$

By subtraction we obtain from these equations

$$H_0 - E_0 - (H_1 - E_1) = L\left(\frac{1}{\sqrt{1 - \frac{v^2}{c^2}}} - 1\right). \qquad [(7)]$$

The two differences of the form $H - E$ occurring in this expression have simple physical significations. $H$ and $E$ are energy values of the same body referred to two systems of co-ordinates which are in motion relatively to each other, the body being at rest in one of the two systems (system ($x$, $y$, $z$)). Thus it is clear that the difference $H - E$ can differ from the kinetic energy $K$ of the body, with respect to the other system ($\xi$, $\eta$, $\zeta$), only by an additive constant $C$, which depends on the choice of the arbitrary additive constants of the energies $H$ and $E$. Thus we may place

$$H_0 - E_0 = K_0 + C \qquad [(8)]$$

$$H_1 - E_1 = K_1 + C \qquad [(9)]$$

since $C$ does not change during the emission of light." [5]

Equations (8) and (9) are the key in Einstein's derivation, which is equivalent to a statement that (the change in) non-kinetic energy has the same value in all reference frames, i.e. the difference in energy values of an object measured in two reference frames is only the difference in its values of kinetic energy. This assertion by Einstein has been a major source of controversy regarding the validity of Einstein's derivation in 1905. Ives [12] and Jammer [13] think that the mass-energy equation is implied by eqs. (8) and (9); without justifying them, Einstein's derivation is invalid. However, the current definition of kinetic energy in relativistic mechanics has implied eqs. (8) and (9), which weakens the objection of Ives and Jammer. From eqs. (8) and (9), Einstein derived an approximate mass-energy equivalence.

"So we have

$$K_0 - K_1 = L\left(\frac{1}{\sqrt{1-\frac{v^2}{c^2}}} - 1\right) \qquad [(10)]$$

The kinetic energy of the body with respect to ($\xi$, $\eta$, $\zeta$) diminishes as a result of the emission of light, and the amount of diminution is independent of the properties of the body. Moreover, the difference $K_0 - K_1$, like the kinetic energy of the electron ( § 10), depends on the velocity.

Neglecting magnitudes of fourth and higher orders we may place

$$K_0 - K_1 = \frac{1}{2}\frac{L}{c^2}v^2." [5] \qquad (11)$$

Equation (10) is a logical consequence of eqs. (8) and (9), which states the difference in the values of an object's kinetic energy measured in one reference frame at two time points (i.e. $K_0 - K_1$) equals the difference between the changes of total energy measured in that frame (i.e. $H_0 - H_1$) and the frame where the object is stationary (i.e. $E_0 - E_1$) at these two time points. The right hand side of eq. (11) is an approximate of the right hand side of eq. (10), which gives an appearance of the classical expression of kinetic energy. From this approximate, Einstein draws the conclusion that "if a body gives off the energy $L$ in the form of radiation, its mass diminishes by $L/c^2$".

The transition from eq. (10) to eq. (11) does show Einstein's ingenuity in dealing with difficult problems in physics, but as a relativistic proof of the mass-energy equation, it lacks sufficient logical rigour.

Firstly, $K_0$ and $K_1$ are obviously relativistic kinetic energy, which would not be equal to $\frac{1}{2}mv^2$ because $K = \frac{1}{2}mv^2$ is a classical formula. If relativistic kinetic energy $K_{rel} \neq \frac{1}{2}mv^2$, we cannot say that $K_{rel} = \frac{1}{2}\frac{L}{c^2}v^2$ implies $L = mc^2$ or $E = mc^2$. At least, we cannot say that $K_{rel} = \frac{1}{2}\frac{L}{c^2}v^2$ implies a precise relationship $L = mc^2$ or $E = mc^2$.

Secondly, the mass-energy relationship from Einstein's derivation seems to be velocity dependent. When $v$ is larger, such as $v = 0.8c$, magnitudes of fourth and higher orders cannot be neglected. So $E = mc^2$ derived implicitly by Einstein in 1905 is only an approximate when $v$ is relatively small, it is not a universal relation applicable to objects at all velocities. Einstein in 1946 acknowledged the imprecision of his mass-energy equation by noting that "It is customary to express the equivalence of mass and energy (though somewhat inexactly) by the formula $E = mc^2$" [1].

Thirdly and more importantly, according to our Proposition 3, mass and energy should be measured in the same reference frame, but in eq. (11) $K_0 - K_1$ and $L$ (hence $L/c^2$) are not measured in the same reference frame. $L$ is the radiation energy measured in the frame where the emitting body is stationary, while $K_0$ and $K_1$ are kinetic energy measured in the frame where the emitting body is moving with velocity $v$. As mass-energy equivalence should not be one in frame $(x, y, z)$ and one in frame $(\xi, \eta, \zeta)$, Einstein's "relativistic" derivation fails to show equivalence between mass and energy measured in the same reference frame.

## 5. Reflection on the definition of relativistic kinetic energy

Einstein's eqs. (8) and (9) are among the main controversial points regarding the validity of Einstein's derivation [12, 13]. The two equations are consistent with classical physics where the difference between the values of an object's energy measured by two reference frames in relative motion is only kinetic energy. Since special relativity also postulates those, we obtain the expression for relativistic kinetic energy from the work done to produce the velocity between two reference frames,

$$K = W = \int_{x_1}^{x_2} F dx = \int_{x_1}^{x_2} \frac{m_0 a dx}{(1 - v_x^2/c^2)^{3/2}} = \int_o^v \frac{m_0 v_x dv_x}{(1 - v_x^2/c^2)^{3/2}} = \frac{m_0 c^2}{\sqrt{1 - v^2/c^2}} - m_0 c^2$$

(12)

In eq. (12), $W$ is work, $F$ force, $x_1$ and $x_2$ the object's positions, $a$ acceleration, $v_x$ velocity in the $x$-axis direction.

The relativistic definition of kinetic energy seems not symmetric with other relativistic quantities. The relativistic momentum is

$$P = m_0 v / \sqrt{1 - v^2/c^2} .$$
(13)

Though physicists cannot agree on whether physics should have the concept of relativistic mass, relativistic mass is

$$m = m_0 / \sqrt{1 - v^2/c^2}. \qquad (14)$$

Laue [23] an Klein [24] have also shown that the relativistic total energy is

$$E = E_0 / \sqrt{1 - v^2/c^2}. \qquad (15)$$

It seems a bit inconsistent that the relativistic kinetic energy and non-kinetic energy do not share such a concise transformation relation as the total energy.

If we postulate that kinetic energy has the same transformation as total energy, kinetic energy would be written in relativistic form as

$$K = \frac{1}{2} \frac{mv^2}{\sqrt{1 - v^2/c^2}}. \qquad (16)$$

Then, the relativistic non-kinetic energy would be

$$E_{non-kinetic} = \frac{m_0 c^2}{\sqrt{1 - v^2/c^2}} - \frac{1}{2} \frac{m_0 v^2}{\sqrt{1 - v^2/c^2}} = \frac{m_0 c^2 - m_0 v^2/2}{\sqrt{1 - v^2/c^2}} \qquad (17)$$

Defining kinetic and non-kinetic energy as such appears to be more consistent with the spirit of special relativity and more symmetric with definitions of other relativistic quantities. Such definitions would invalidate eqs. (8) and (9) and consequently Einstein's derivation in 1905. Einstein's classical derivation in 1946 is not affected by such a change in the definition of relativistic kinetic energy.

## 6. Derivation of mass-energy equation from conservation of momentum

Without eqs. (8) and (9), Einstein could have started with momentum conservation to derive the mass-energy relation. Then in the frame ($x$, $y$, $z$) where the radiating body is at rest, we have

$$P_{S0} = P_{S1} + \frac{E_S}{2c} - \frac{E_S}{2c} = P_{S1} = 0 \tag{18}$$

In eq. (18), *P* stands for momentum, the subscript *S* indicates the frame where the radiating body is stationary, and $\frac{E}{2c}$ is the momentum of light wave packet in one direction (as in Maxwell's classical electromagnetic theory, here Einstein's *L* is replaced with the more conventional *E* for energy).

In the frame ($\xi, \eta, \zeta$) where the radiating body is moving at the velocity *v*,

$$P_{V0} = P_{V1} + \frac{E_S}{2c} \frac{1 + \frac{v}{c}\cos\varphi}{\sqrt{1 - v^2/c^2}} - \frac{E_S}{2c} \frac{1 - \frac{v}{c}\cos\varphi}{\sqrt{1 - v^2/c^2}} = P_{V1} + E_S \frac{\frac{v}{c^2}\cos\varphi}{\sqrt{1 - v^2/c^2}} \tag{19}$$

In eq. (19), the subscript *V* indicates the moving frame. When $\phi = 0$,

$$\Delta P_V = P_{V0} - P_{V1} = \frac{\frac{v}{c^2}E_S}{\sqrt{1 - v^2/c^2}} \tag{20}$$

Since $P_V = m_V v = m_S v / \sqrt{1 - v^2/c^2}$ (here relativistic mass $m_V$ is used for illustration purpose),

$$\Delta m_S = E_S / c^2. \tag{21}$$

In the frame where the radiating body is stationary, when energy *E* is emitted, there is a loss of mass $\Delta m = E/c^2$. This mass-energy equivalence in the same reference frame is exact rather than approximate, which has been confirmed by experiments.

From eq. (20) and $P_V = m_V v$, we can also obtain

$$\Delta m_V c^2 = E_S / \sqrt{1 - v^2/c^2}$$

Since $\Delta m_S c^2 = E_S$, let $\Delta m_V c^2 = E_V$, which is the energy (value) measured in the frame moving relative to the radiating body, we obtain

$$E_V = E_S / \sqrt{1-v^2/c^2} = \Delta m_S c^2 / \sqrt{1-v^2/c^2} \ . \tag{22}$$

Equation (22) is the relativistic formula describing the relationship between values of the same energy measured in two reference frames, which depends on their relative velocity *v*.

If we use subscript 0 to indicate measurements obtained in the frame where the radiating body is stationary, our new derivation reveals what Einstein should have proved is eq. (15) derived by Laue [23] and Klein [24]

$$E = E_0 / \sqrt{1-v^2/c^2} \ .$$

Equation (15) corresponds to the relativistic mass equation [25]

$$m = m_0 / \sqrt{1-v^2/c^2} \ .$$

The essence of Einstein's derivation in 1905 is actually an approximation of eq. (15),

$$E - E_0 = \frac{E_0}{\sqrt{1-v^2/c^2}} - E_0 = E_0 (\frac{1}{2}\frac{v^2}{c^2} + \frac{3}{8}\frac{v^4}{c^4} + \frac{5}{16}\frac{v^6}{c^6} + \cdots) \approx \frac{1}{2}\frac{E_0}{c^2}v^2 \ . \tag{23}$$

Expanding the relativistic mass equation and using classical kinetic energy expression $K = \frac{1}{2}mv^2$ can get the same relationship when *v* is small,

$$\Delta m = m - m_0 = \frac{m_0}{\sqrt{1-v^2/c^2}} - m_0 = m_0 (\frac{1}{2}\frac{v^2}{c^2} + \frac{3}{8}\frac{v^4}{c^4} + \frac{5}{16}\frac{v^6}{c^6} + \cdots) \approx \frac{1}{2}\frac{m_0 v^2}{c^2} = \frac{E_0}{c^2} \tag{24}$$

However, both eq. (24) and Einstein's derivation in 1905 describe relationships between variables measured in different frames, which violate Proposition 3, and need classical kinetic energy formula, which violates Proposition 2.

Therefore, the relativistic result should be $E = E_0 / \sqrt{1-v^2/c^2}$, which is just a different expression of the relativistic mass equation $m = m_0 / \sqrt{1-v^2/c^2}$. This relationship between energy values measured in two reference frames has been shown by Laue, using conservation of energy-momentum tensor and assuming that there is no energy flow in the rest frame [23]. Klein extended Laue's results to closed system with or without flow of energy [24].

## 7. Shortcomings in Einstein's derivation in 1946 and correct derivation using Einstein's premise

Einstein's derivation in 1946 has the shortcoming of not distinguishing different values measured in the two reference frames. A wave complex has different energy values in two frames $K_0$ and $K$ with relative motion. In eqs. (1) and (2), the energy values of the wave complexes are those measured in frame $K_0$, while the momentums are measured in frame $K$. The derivation is logically inconsistent, because mass-energy equivalence should be the equivalence when both mass and energy are measured in the same reference frame.

To derive a more precise mass-energy equation, we need to know in which reference frame the variables are measured. We can firstly add subscripts to the variables so that we can keep track of the reference frames in which they are measured. We re-write eqs. (2a) and (2b) as

$$m_{K1} v + v \cdot E_K / c^2 = m_{K2} v. \qquad (25a)$$

$$\Delta m_K = m_{K2} - m_{K1} = E_K / c^2. \qquad (25b)$$

In eqs. (25), $m_{K1}$ is the mass before the absorption in the moving frame, $m_{K2}$ the mass after the absorption in the moving frame, and $E_K$ the energy measured in the moving frame. From eq. (25b), we obtain the mass-energy equation in the moving frame

$$E_K = \Delta m_K c^2. \qquad (25c)$$

So far, the derivation is in classical physics with electromagnetic waves having momentum. What is the relationship between mass and energy in the stationary frame?

Lorentz relativistic mass formula has given us the relationship between values of a mass in different reference frames. Using Lorentz relativistic mass formula, we obtain from eq. (25b)

$$\frac{\Delta m_0}{\sqrt{1-v^2/c^2}} = \frac{m_{02}}{\sqrt{1-v^2/c^2}} - \frac{m_{01}}{\sqrt{1-v^2/c^2}} = \frac{E_K}{c^2},$$

which gives

$$E_K = \Delta m_0 c^2 / \sqrt{1-v^2/c^2}. \tag{26}$$

When $v = 0$, we have the mass-energy equation in the stationary frame

$$E_0 = \Delta m_0 c^2. \tag{27}$$

Therefore, the relativistic energy formula is still eq. (15), i.e., what Laue [23] and Klein [24] have found

$$E = E_0 / \sqrt{1-v^2/c^2}.$$

Equation (15) is the correct formula for relationship of relativistic energy values between two reference frames with relative motion. The result reveals the symmetry between changes in relativistic mass and in relativistic energy in the moving frame. The equation $E = mc^2$ can be obtained approximately from the correct relativistic equation only when classical kinetic definition $K = \frac{1}{2}mv^2$ is used and the requirement of measuring mass and energy in the same reference frame (Proposition 3) is not stuck to.

## 8. Definition of momentum and the mass-energy relation

Strictly speaking, the two derivations presented in this paper and many other derivations so far are only illustrations of the mass-energy equivalence contained in Newtonian mechanics and Maxwell's electromagnetic theory with special scenarios. Einstein in 1935 tried to prove rest energy $E_0 = m$ by asserting without proof that total energy $E = E_0 + m\left(\frac{1}{\sqrt{1-v^2/c^2}} - 1\right)$ and kinetic energy is $m\left(\frac{1}{\sqrt{1-v^2/c^2}} - 1\right)$. However, he did not give a derivation of $E_0 = \Delta m_0 c^2$ [26]. Since in Newtonian mechanics $m = P/v$, Maxwell's electromagnetic momentum $P = E/c$ implies

$$m = \frac{P}{v} = \frac{E/c}{c} = \frac{E}{c^2}. \tag{28}$$

If in Newtonian mechanics there were another type of momentum which had no corresponding mass or inertia, $P_{WithoutMass} \neq mv$, eqs. (2), (21) and (25) and all other similar equations would not be valid. If $m \equiv P/v$ or $P \equiv mv$, we can obtain the mass-energy equation directly from $P \equiv mv$ and electromagnetic momentum $P = E/c$. When a material object with mass $m$ is converted completely into electromagnetic waves, the total energy released is $E = mc^2$. The mass-energy equation can be derived from classical physics without involving special relativity, while in Einstein's "relativistic" derivation $E = mc^2$ cannot be obtained without resorting to classical kinetic energy definition and approximation at small velocity. Even with classical kinetic energy definition and approximation at small velocity, Einstein still failed to prove $E = mc^2$ for mass and energy measured in the same reference frame.

## 9. Discussion on some incorrect views

During the process of communicating the results of this study with researchers in this field, some incorrect views on the mass-energy equation and results of this study emerge. The following three views are representative of these incorrect views.

First, some researchers thought that $E = mc^2$ can be derived only when the constancy of the speed of light is postulated. This view is obviously ignorant of the history of physics. Preston [27]; Poincaré [19], De Pretto [28] and Hasenöhrl [29] had proposed similar mass-energy relations well before Einstein postulated the constancy of the speed of light. The speed of light c is the constant in Maxwell's electromagnetic equations, which is the velocity of light in its medium. As mass does not change in class physics, the corresponding energy contained in the rest mass is also dependent on the constant velocity of light in its medium. We might say that classical physics cannot derive the relation $E = E_0 / \sqrt{1 - v^2/c^2}$, but the assertion that $E = mc^2$ can be derived only when the constancy of the speed of light is postulated is obviously wrong. Even Einstein [22] and Lewis [20] derived the mass-energy relation without resorting to the constancy of the speed of light or special relativity.

Second, some researchers thought that $E = E_0 / \sqrt{1 - v^2/c^2}$ has been known to physicists for a long time, there is no new finding in arguing whether $E = mc^2$ can be derived from classical physics. Given that $E = mc^2$ being a relativistic result has become a universal belief in modern society, establishing the true identity of $E = mc^2$ is not only important in physics, but also significant in philosophy and history of science.

Third, some researchers thought that derivations in sections 6 and 7 used similar assumptions as Einstein, so that $E = mc^2$ cannot be considered as a result of classical physics as well. Derivations in sections 6 and 7 are intended to illustrate the relationship between $E = mc^2$ and $E = E_0 / \sqrt{1 - v^2/c^2}$ within the framework of special relativity; of course the relativistic assumptions should be used. This does not affect the fact that $E = mc^2$ can be derived from classical physics.

## 10. Conclusions

From the preceding analysis, we may draw the following conclusions:

Firstly, the mass-energy equation $E = mc^2$ is contained in Maxwell's classical electromagnetic theory and the momentum definition of Newtonian mechanics. With the momentum definition in Newtonian mechanics $P \equiv mv$ and Maxwell's electromagnetic momentum $P = E/c$, the mass-energy equation $E = mc^2$ should be a logical consequence.

Secondly, all logically valid derivations of $E = mc^2$, where both mass $m$ and energy $E$ are measured in the same reference frame, rely on the two classical equations $P \equiv mv$ and $P = E/c$. No matter whether a derivation is under classical or relativistic conditions, the two equations must be held true. If the two equations are denied in any of those derivations, it is not possible to arrive at $E = mc^2$ logically. If these two equations are held true, the mass energy equation $E = mc^2$ can be obtained directly without the special scenarios assumed for those derivations.

Thirdly, since $E = mc^2$ can be derived without resorting to any relativistic result, it is a formula from classical physics, applicable to both classical physics and special relativity when relativistic mass is used in the equation.

Fourthly, the relativistic transformation of energy (values) between different reference frames is $E = E_0 / \sqrt{1 - v^2/c^2}$.

Fifthly, Einstein's "relativistic" derivation in 1905 relies on classical kinetic energy definition, describes implicitly a relationship between mass and energy measured in different reference frames and leads only to an approximation at low velocity for a velocity dependent equation; hence it is not logically valid as a relativistic proof of the mass-energy equation $E = mc^2$.